\documentclass{osa-article}

\journal{oe}

\usepackage{lineno}
\usepackage{multirow}
\usepackage{graphicx}

\begin{document}

\title{Group frame neural network of moving object ghost imaging combined with frame merging algorithm}

\author{Da Chen,\authormark{1,2} Shan-Guo Feng,\authormark{2} Hua-Hua Wang,\authormark{2} Jia-Ning Cao,\authormark{1} Zhi-Wei Zhang,\authormark{1} Zhi-Xin Yang,\authormark{2} Ao Yan,\authormark{2} Lu Gao,\authormark{2,*} and Ze Zhang,\authormark{3,4,*}}

\address{\authormark{1}School of Information Engineering, China University of Geosciences, Beijing 100083, China\\
\authormark{2}School of Science, China University of Geosciences, Beijing 100083, China\\
\authormark{3}Qilu Aerospace Information Research Institute, Jinan 250100, China\\
\authormark{4}Aerospace Information Research Institute, China Academy of Sciences, Beijing, China}

\email{\authormark{*}gaolu@cugb.edu.cn, zhangze@airas.ac.cn}



\begin{abstract}
The nature of multiple samples to extract correlation information limits the applications of ghost imaging of moving objects. A novel multi-to-one neural network is proposed and the concept of "batch frame" is introduced to improve the serial imaging method. The neural network extracts more correlation information from a small number of samples, thus reducing the sampling ratio of the ghost imaging technique. We combine the correlation characteristics between images to propose a frame merging algorithm, which eliminates the dynamic blur of high-speed moving objects and further improves the reconstruction quality of moving object images at a low sampling ratio. The experimental results are consistent with the simulation results.
\end{abstract}

\section{Introduction}
Ghost imaging (GI) is a new type of imaging technology\cite{1,2}. It is different from the previous measurement of the light intensity distribution on the surface of an object to obtain an image. It is an active imaging technique that acquires object information based on the higher-order correlation of the light field. GI allows for lens-free imaging since no lenses are required\cite{3}. Thus it is possible to broaden to some special wavebands, such as x-ray\cite{4,5,6,7,8}, far and near-infrared\cite{9,10}, and terahertz waves\cite{11,12,13}.\par
It often involves moving objects when ghost imaging is put into practical applications, such as security surveillance, blood cell classification counts\cite{14}, and military long-range target detection. The characteristics of GI make it necessary to measure multiple times to obtain an image of high quality. The motion of the object can cause blur, which reduces the image quality. Therefore, the critical issues facing the applications of GI to moving objects are how to improve the imaging speed of GI and maintain high image quality.  \par
Many works focus on improving the imaging speed of moving objects by increasing the refresh frequency of the light source\cite{15} or improving the imaging algorithm. In 2018, Ming-Jie Sun et al. developed an LED array light source with a refresh frequency of up to 500 kHz and achieved 1000 Hz imaging of simple scenes\cite{16}. The following year, Wei-Gang Zhao et al. increased the refresh frequency of the $10\times10$ LED light source array to 100 MHz and achieved fast imaging at 1.4 MHZ in the laboratory\cite{17}. In addition, some study has attempted to improve imaging algorithms, such as using a priori information about moving objects in the scene such as velocity\cite{18,19,20}, position\cite{21,22}, and sparsity\cite{14} to improve the imaging speed and image quality. In 2012, Cong Zhang’s research group achieved the imaging of moving objects by using displacement compensation for the reference beam\cite{23}. In 2019, Wei-Tao Liu et al. proposed the imaging algorithm of cross-correlation. The displacement of the object is obtained by calculating the correlation of the image, and a clear image of the object is gradually reconstructed during the object's motion\cite{24}. With the widespread applications of Deep Learning (DL) in areas like image noise reduction, image restoration, and natural language processing\cite{25,26,27}, DL has also been widely applied to computational imaging, such as scattering medium and turbid medium imaging\cite{28,29,30,31}, lensless imaging\cite{32}, and ghost imaging\cite{33,34,35,36}. In 2020, Wei-Tao Liu et al. used the convolutional denoising auto-encoder (CDAE) to improve the imaging quality of moving objects\cite{37}.\par
Based on deep learning, we propose a new network and introduce the concept of "batch group frame", which effectively solves the problems of slow imaging speed and high samples of GI. At a low sampling ratio of 3.125\%, the SSIM index is 26 times higher than that of GI, and the PSNR index is 3 times higher than that of GI. In addition, we propose the frame merging algorithm (FMA) considering the correlative of the images at different positions, which effectively eliminates the motion blur of moving objects with high-speed rotation. The frame images generated by the algorithm are put into the network for training, which improves the quality of moving object ghost imaging and reduces the overall sampling ratio. In the experiment of Gl for moving objects with high rotation speed, high-quality images can be achieved with SSIM of 0.893 and PSNR of 21.97. This approach provides a new method for the recovery of moving objects at a low sampling ratio.\par

\section{Methods and network}
\subsection{Multi-to-one Group Frame Neural Network (GFNN) structure}
Conventional ghost imaging systems consist of two spatially separated beams, one for the reference beam and the other for the object beam. The reference beam does not interact with the object to be measured and propagates freely to a high-spatial-resolution detector. The detector picks up the spatial distribution of its light intensity. The object beam interacts with the object and is collected by bucket measurement to obtain a 1-dimensional intensity signal. The object function is represented as $T(x, y)$, where x and y are the horizontal and vertical coordinates in the object image plane. The object is illuminated by a set of speckle patterns $I_m(x, y)$, and the subscript integer $(m = 1,2,3,..., M)$ denotes the $Mth$ speckle pattern. Thus, the intensity signal sequence $S_m$ (bucket measurements) collected by the bucket detector is expressed as
\begin{equation}\label{eq1}
S_m=\iint T(x,y)I_m(x,y)\mathrm{d}x\mathrm{d}y,
\end{equation}
The image of the object is to be recovered by correlating the intensity of the light field between the two optical signals denoted as Eq. (\ref{eq2}).
\begin{equation}\label{eq2}
T_{GI}(x,y) = \langle \sum_{i=1}^m S_iI_i(x,y)\rangle - \langle \sum_{i=1}^m S_i\rangle \langle \sum_{i=1}^m I_i(x,y)\rangle,
\end{equation}
where $\langle \cdot \rangle$ denotes the averaging operation. $S_m$ is the bucket measurement and $I_m (x,y)$ is the speckle pattern.\par
\begin{figure}[htbp]
\centering
\includegraphics[width=0.78\linewidth]{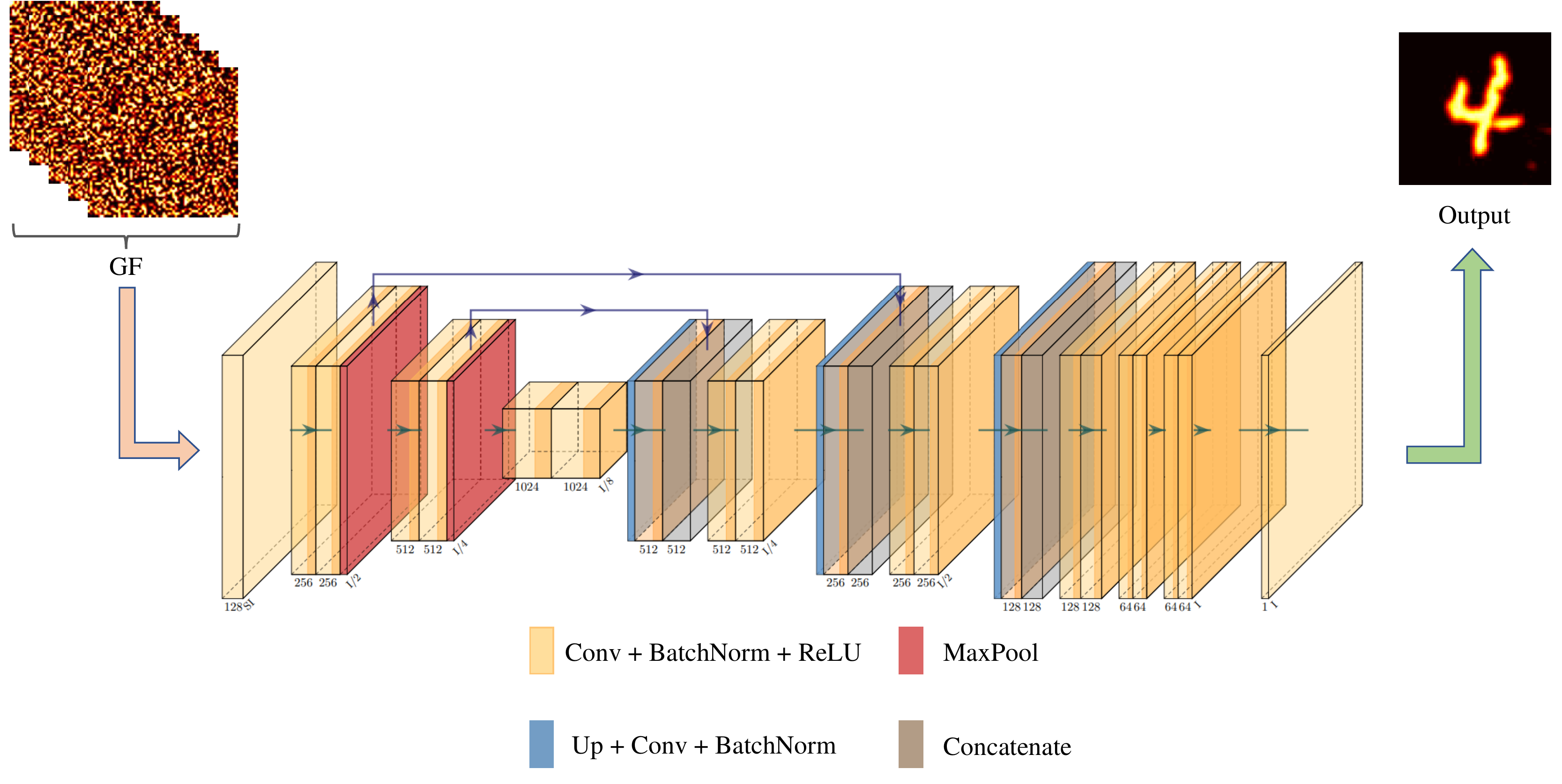}
\caption{The network structure of GFNN. It takes GF as input and outputs a single image.}
\label{fig1}
\end{figure}
The core problem limiting the development of GI is still the excessive number of samples, resulting in a long imaging time. In methods that use deep learning GI (GIDL) to reduce the sampling ratio, the target image is often generated in a one-to-one style. For example, the blurry images are matched one-to-one with their corresponding images of the object\cite{37}, or the GI images are matched one-to-one with their corresponding images of the object\cite{38}. Here, we propose a special network which is shown in Fig. \ref{fig1}. Its input is the image after multiplying the bucket measurements with the speckle pattern, which we name the bucket measurement image. Since in the generation of GI, if the number of samples is $m$, there will be $m$ frames of bucket measurement images $m\ast S_iI_i(x,y)$ corresponding to it, so the input to the network is of multi-to-one type.The specific inputs are represented as follows
\begin{equation}\label{eq3}
GF(x_i,y_i,i)= (S_iI_i(x,y),i),
\end{equation}
The input of the network is defined as Group Frame (GF). $S_iI_i(x,y)$ denotes the bucket measurement image. Here we have $i = 1,2,\cdots,m$, enumerating the total $m$ samples. $GF(x,y,m)$ denotes the 3-dimensional matrix formed by a set of bucket measurement images $SI(x,y)$ with $m$ frames sampled.\par
Then the input/output relationship of this network can be expressed as
\begin{equation}\label{eq4}
O(x,y)=F\{GF(x,y,m)\},
\end{equation}
where $O(x,y)$ represents the output of the network, and $F\{\cdot \}$ represents the network that fits the GF to the predicted image. \par
The training process of the network is represented as
\begin{equation}\label{eq5}
F = arg \min_{\theta \epsilon \Theta}\sum_{i=1}^N L(T_i(x,y),F_{\theta}\{ GF_i(x,y,m)\}) + W(\theta),
\end{equation}
where $L(\cdot )$ is the loss function used to guide the network fitting process, which quantifies the difference between the network output and the ground truth. $\Theta$ is the total parameters of the network. The subscript $i$ denotes the $ith$ input/output, which takes values in the range $i = 1,2,\cdots,N$. In addition, $w(\theta)$ is added to regularize parameters in case of network overfitting.\par
The Group Frame Neural Network (GFNN) is a variation of the U-net\cite{39}, which replaces the first two layers of the downsampling part with GF and reduces the number of channels by a multiple convolution operation after upsampling. The network retains the rest of the U-net: a jump connection layer is used to connect the upper and lower sampling layers, sharing parameters and information, which prevents degradation of the training effect and overfitting problems. Finally, the image is output by a $1\times1$ convolution operation. GF extracts the features in the network through the downsampling layer, scales up the image size in the upsampling layer, and outputs the predictions after preventing degradation in the jump connection layer.\par
\begin{figure}[htbp]
\centering
\includegraphics[width=0.8\linewidth]{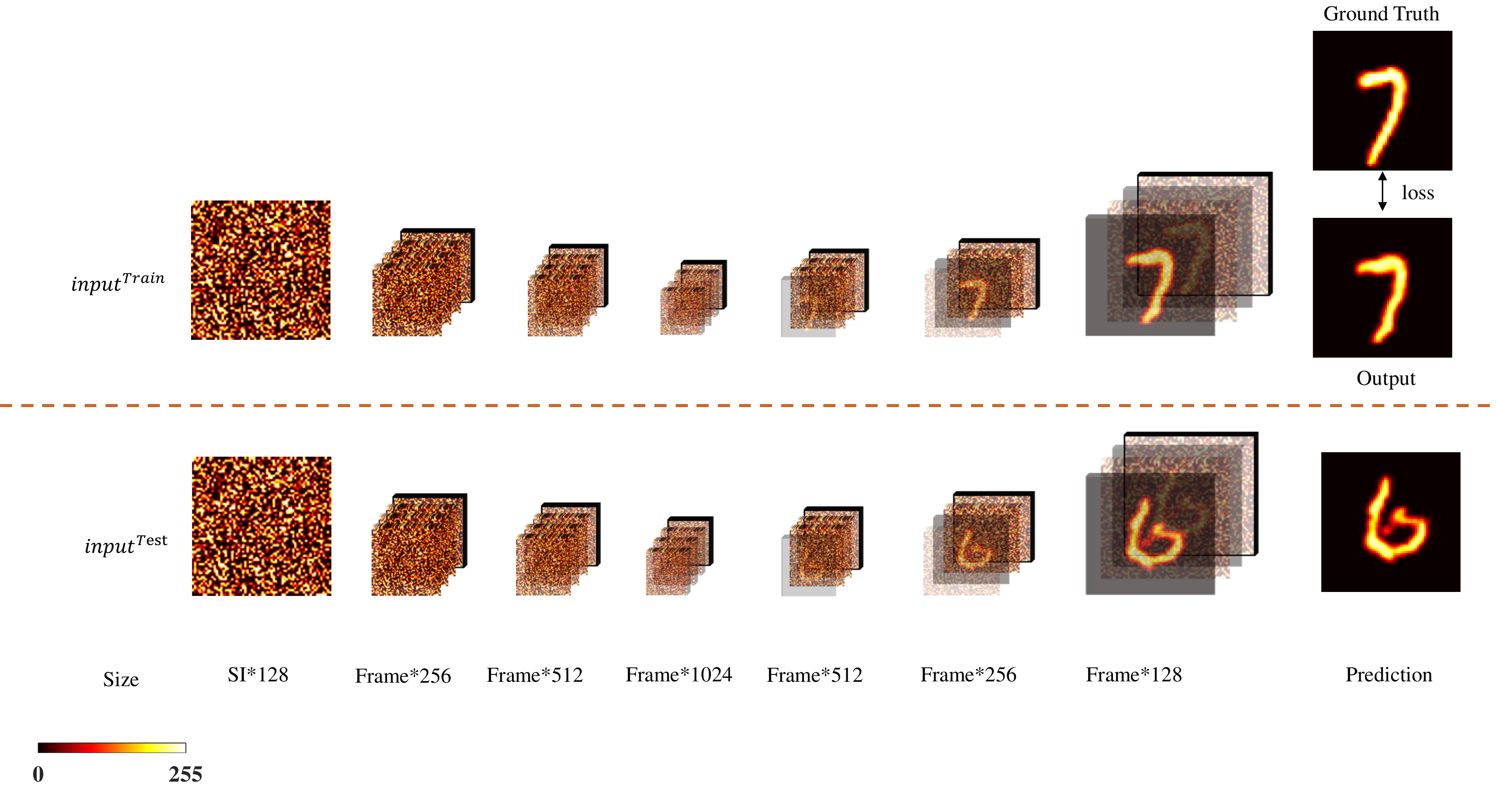}
\caption{The training and testing process of GFNN. The network extracts the information contained in the GF and transforms it into a fixed-length vector. After changing the size, it is transformed into an output sequence. It can be seen that the GF gradually fits towards the real image, which shows that the network can extract the invariant object information in the GF.}
\label{fig2}
\end{figure}
Since the nature of upsampling and downsampling operations is to compress information and extract features, it is not possible to obtain more information than the network input. 
Compared with the GI obtained through Eq. (\ref{eq2}), the GF contains all the information of the images, and thus inputting the GF into the network instead of the GI can provide more optional directions for the training of the network and make the network have more trends in the fitting process. Figure \ref{fig2} shows the invariant features of the images extracted from the network layer by layer.\par
GF is $n$ times larger than GI at $n$ samples, making the training time of the network significantly longer. To reduce the network input/output overhead while maintaining its high generalization ability\cite{40}, we provide an improvement to the traditional serial ghost imaging method.
By referring to the work of T. Bian et al\cite{41}, we introduce the concept of "batches" in deep learning. A "batch group frame (BGF)" is defined as a collection of multiple GFs. The first dimension of BGF represents the number of images in each batch, the second and third dimensions represent the two dimensional size of the images, and the fourth dimension represents the number of frames of each image. In the training process, the 4-dimensional array of BGF is considered as the smallest operation unit, and the BGF can be operated in parallel with the help of the Graphics Processing Unit (GPU). The method can process 128 times more data than traditional serial imaging algorithms, and the imaging speed is 70 times faster. The data is measured by a computer timer.\par
Conventional computational ghost imaging takes the same set of speckle patterns for higher imaging speed, which makes the generalization ability of the network poor\cite{40}. Therefore, the method of using the same set of speckle patterns within the same BGF and different sets of speckle patterns between different BGFs is chosen\cite{41}. The speckle patterns are generated by a spatial light modulator (SLM, HOLOEYE, PLUTO-2-VIS-016).\par
A computer with a GTX A6000 graphics card is used to train the GFNN network with 2000 images from the MNIST datasets and 20000 images from the Fashion-MNIST datasets. The size of the images is $64\times64$ pixels and the sampling ratio is 3.125\%. The training sessions take 22 and 12 minutes for the above datasets. In the test session, GFNN spends 6 $ms$ to process a single image.\\
\subsection{Frame merging algorithm (FMA)}
Ghost imaging requires multiple measurements to reconstruct the image, which needs a long imaging time. The movement of the object will lead to blurring images. Based on the GFNN network proposed in this study, we improve the existing motion-imaging algorithm.\par
Existing work demonstrates that images can be recovered by tracking compensation\cite{42} or by obtaining the trail of the object via the cross-correlation between sequential low-sampling GIs\cite{24}, on the principle that acquiring the object position requires much fewer samples than obtaining a clear object image. Thus the core of reconstructing a clear moving object image lies in acquiring the displacement of the object.\par
For an object rotates at a constant speed, the cross-correlation\cite{24} of the GI images at two positions can be expressed as\par
\begin{equation}\label{eq6}
CCG(\Delta \theta) = \sum_{m,n}G_{\theta}(m,n)\circ G_{\theta-\Delta \theta}(m,n),
\end{equation}
where $G_{\theta}(m,n)$ and $G_{\theta-\Delta \theta}(m,n)$ present the two GI with the normalization operation, and $\circ$ represents the Hadamard product of two matrices, which means multiplying each element of the matrix correspondingly.\par
The information of the moving object itself is invariant during the rotation. Thus for objects at two locations, the information about themselves is interdependent, while the noise is statistically independent\cite{24}.
Therefore, the correlation is maximum when $CCG(\Delta \theta)$ obtains the maximum value. $\Delta \theta$ is the real angle $\theta_{real}$ between two positions during the uniform rotation of the object, which can be expressed as $\theta_{real} = argmax(CCG(\Delta \theta))$. Then $G_{\theta}(m,n)$ or $G_{\theta-\Delta \theta}(m,n)$ is superimposed after keeping one stationary and the other rotating $\Delta \theta$ degrees relative to each other. Since the noise of the images is independent, continuously superimposing the images can suppress the noise and reconstruct high-quality images in motion.\par
The correlation algorithm is based on the achieved sequential low-sampling GIs. These GIs are achieved when the object is considered stationary. For example, the limit of the samples for a single image with lateral motion\cite{43} is
\begin{equation}\label{eq7}
N \leq \frac{f}{w} \theta_r,
\end{equation}
the sampling frequency is $f$, the angular resolution of the imaging system is $\theta_r$, the angular velocity of the object motion is $w$, and the number of samples is $N$.\par
When the object is moving at a high speed, the samples per GI are very limited. Using Eq. (\ref{eq6}) to obtain correlation information between images will introduce errors due to the weak correlation between low-sampling GIs. In contrast, constructing a low-sampling GI with more than the upper limit of Eq. (\ref{eq7}) samples will produce motion blur, resulting in lower image quality. Therefore, we extend Eq. (\ref{eq6}) to the frame level, lending the GFNN network for processing.\par
Define the correlation at the frame level as follows
\begin{equation}\label{eq8}
CCF(\Delta \alpha) = \sum_{m,n}F_{\theta}(m,n)\circ F_{\theta-\Delta \theta}(m,n),
\end{equation}
where $F_{\theta}(m,n)$ and $F_{\theta-\Delta \theta}(m,n)$ represent two normalized images within the same batch group of frames (BGF) but between different groups of frames (GF). Since the same group of speckle patterns is taken within the same BGF, the corresponding rotation angle $\Delta \alpha$ between frames within the same BGF but different GF can be calculated by Eq. (\ref{eq8}). And Eq. (\ref{eq8}) is processed based on each frame image but not average images of multiple frames. so that it appears obvious advantages of removing blur. The average rotation angle between two GF is obtained by taking the average value. Then divide by the number of frames $v$ of the GF to get the rotation angle $\alpha$ between two adjacent frames within the GF.\par
\begin{equation}\label{eq9}
\alpha = \frac{1}{v^2}\sum_{i=1}^v argmax(CCF_{(m+i)(n+i)}(\Delta \alpha)),
\end{equation}
where the number of frames between $m$ and $n$ is $v$. The derived $\alpha$ is the rotation angle between two adjacent frames. Since the object rotates at a uniform speed, the rotation angle between BGF can be converted.\par
\begin{figure}[htbp]
\centering
\includegraphics[width=0.7\linewidth]{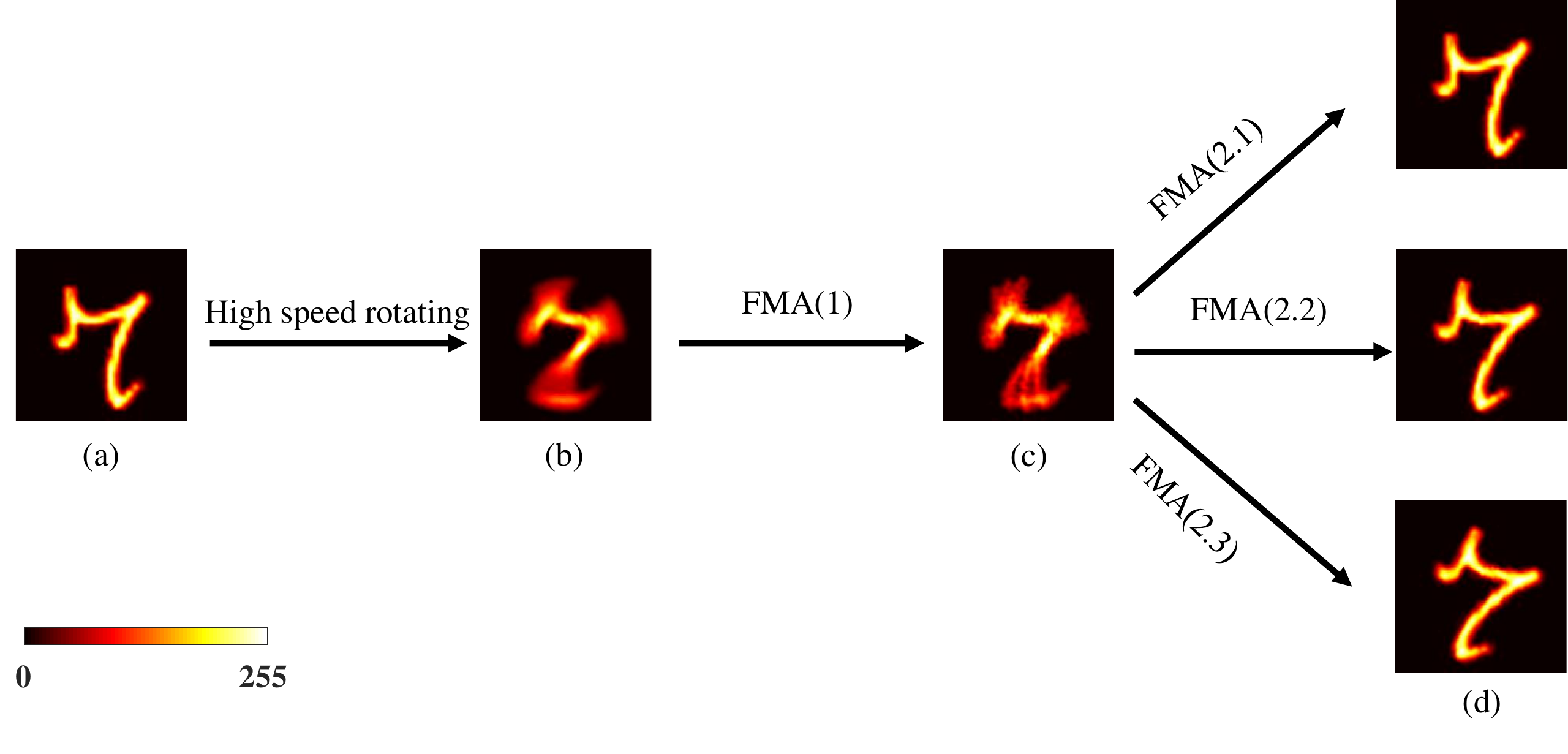}
\caption{FMA algorithm process. The number of batches is set to 3.}
\label{fig3}
\end{figure}

The flow of the Frame Merging Algorithm (FMA) to eliminate the blurring of moving objects is shown in Fig. \ref{fig3}, which consists of the following two steps. FMA(1): Use Eq. (\ref{eq9}) to calculate $\alpha$ between the two adjacent frames and merge BGF in the same position. FMA(2): Get the angle $v\alpha$ between different BGFs and make different BGFs merge in the same position. (Using the location of different BGFs as a base leads to different merging position)\par
Taking the number "7" as the detected object, we superimpose part of the trajectory of the object as shown in Fig. \ref{fig3}(b). The algorithm shows three clear outlines of the image by FMA(1) as shown in Fig. \ref{fig3}(c). The different base positions set by FMA(2) restore different locations, as shown in Fig. \ref{fig3}(d). ( The images produced by the algorithm are all cluttered frame images. To better demonstrate the algorithm flow, we additionally use Eq. (\ref{eq2}) to process the frame images, and the results are as shown in Figs. \ref{fig3}(b), \ref{fig3}(c), and \ref{fig3}(d) above.)\par
The algorithm FMA is only limited to the CCD sampling speed, and the current limit of the detector is $80ms$ for each frame. Our scenario contains two processing. The first processing is to use the algorithm FMA to remove the motion blur of the image, and the second processing is to denoise the image separated by FMA based on the proposed GFNN network. The process is as follows: Frames of moving objects acquired in the ghost imaging system are merged to the same location by FMA to form a GF. Then the GF is used as the input to the GFNN network for training to recover the object images. Since the same set of speckle patterns is used within the BGF, combining them does not reduce the statistical noise. Only the different sets of speckle patterns used between BGFs have the effect of suppressing the statistical noise.\par

\section{Results and analysis}
The training sets chosen for the simulation are 2000 and 20000 images from the MNIST and Fashion-MNIST datasets. The size of the images is $64\times64$ pixels and the sampling ratio is 3.125\%. The two datasets are trained for 100 and 5 epochs, respectively. SSIM and PSNR\cite{44,45,46} are selected as two metrics to evaluate the image quality.\par
\begin{equation}\label{eq10}
PSNR = 10 \cdot lg \left[\frac{255^2}{\frac{1}{MN}\sum_{i=1}^M\sum_{j=1}^N \left[G_0(i,j)-G(i,j)\right]^2}\right],
\end{equation}
where the larger PSNR indicates the better quality of the image. $G_0(i,j)$ is the real image and $G(i,j)$ is the output image. They both contain $i\times j$ pixels.\par
\begin{figure}[htbp]
\centering
\includegraphics[width=0.60\linewidth]{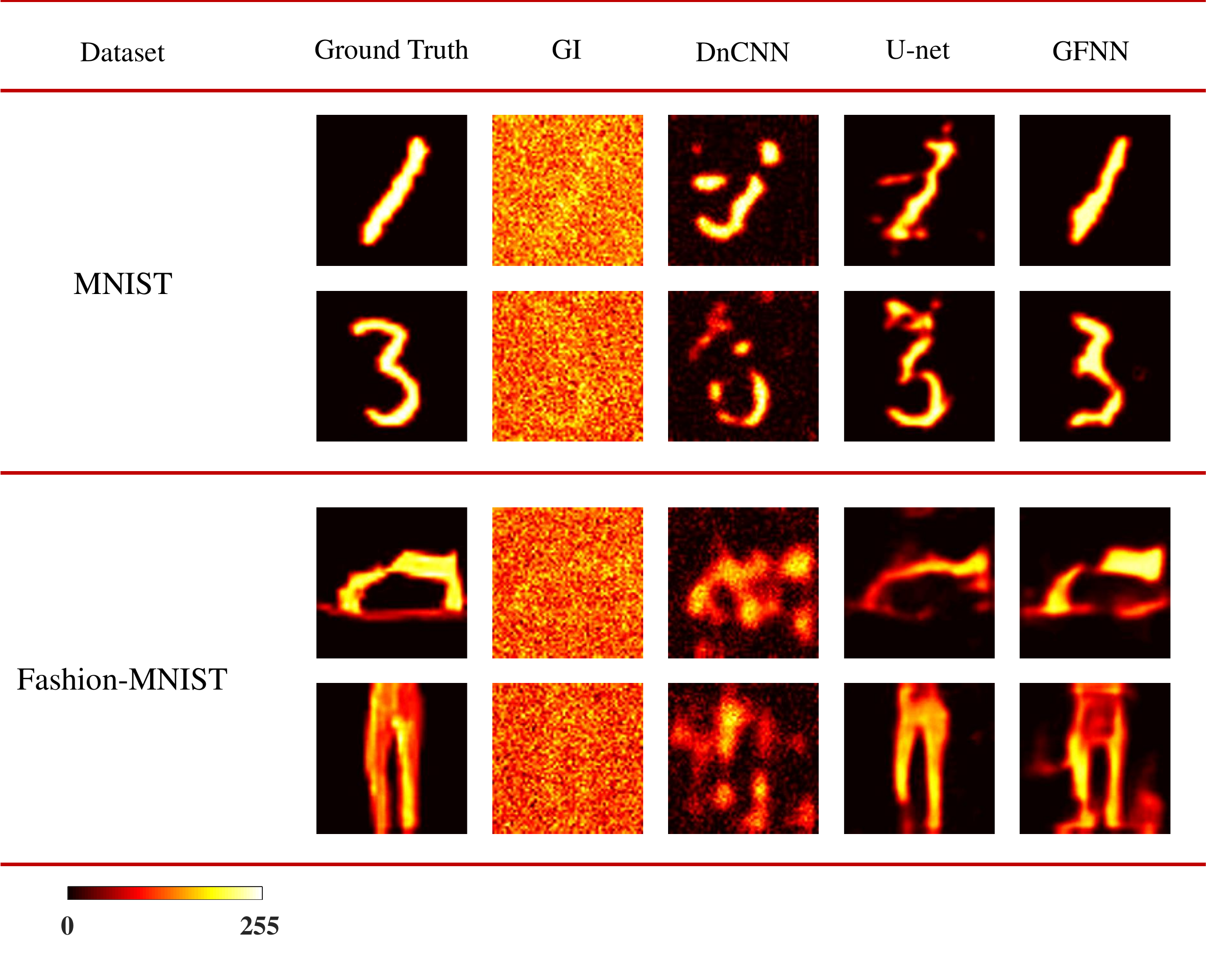}
\caption{The simulation results of GI, DnCNN, U-net, and GFNN networks.}
\label{fig4}
\end{figure}
\begin{table}[htbp]
\centering
\caption{ SSIM and PSNR values of GI, DnCNN, U-net and GFNN}
\label{tab1}
\begin{tabular}{cccccc}
\hline
    & Images & Digital 1 & Digital 3 & Shoes & Pants\\
\hline
\multirow{4}{*}{SSIM} &  GI & 0.0179 & 0.0227 & 0.0347 & 0.0415\\
& DnCNN & 0.1913 & 0.1744 & 0.1331 & 0.1287\\
& U-net & 0.7656 & 0.7145 & 0.5425 & 0.6919\\
& GFNN & 0.9078 & 0.8266 &0.6165 & 0.5094\\
\hline
\multirow{4}{*}{PSNR} &  GI & 5.4010 & 6.1113 & 6.8837 & 7.9374\\
& DnCNN & 14.7507 & 14.0495 &13.3249 & 13.4435\\
&U-net & 18.9016 & 15.1698 & 15.3824 & 19.6904\\
& GFNN & 21.3522 & 18.5400 & 16.6540 & 16.6203\\
\hline
\end{tabular}
\end{table}
The imaging results of different networks\cite{39,47,48} at a low sampling ratio of 3.125\% are shown in Fig. \ref{fig4}. For the $64\times64$ pixels image, the GI is almost indistinguishable. DnCNN does not perform well at this low sampling ratio. The improvement of SSIM and PSNR is not significant. The GFNN output image is clear and of high quality. Moreover, taking the number 3 as the detected object, since its upper part is almost invisible in GI, none of the other networks predicted a result similar to the ground truth, but GFNN restores that part of the information accurately.\par
The SSIM and PSNR values of GI, DnCNN, U-net and GFNN are shown in Table \ref{tab1}. GFNN achieves significant improvements in both SSIM and PSNR on the MNIST dataset. The output image averages 26 times that of GI in SSIM and nearly 3 times that of GI in PSNR. However, in the predictions of the Fashion-MNIST dataset, U-net and GFNN show similar levels. This demonstrates that GFNN can obtain sufficient information through GF when the amount of data is small, but when the amount of data is too much, GFNN will show a trend of declining training effect.\par
\begin{figure}[htbp]
\centering
\includegraphics[width=0.55\linewidth]{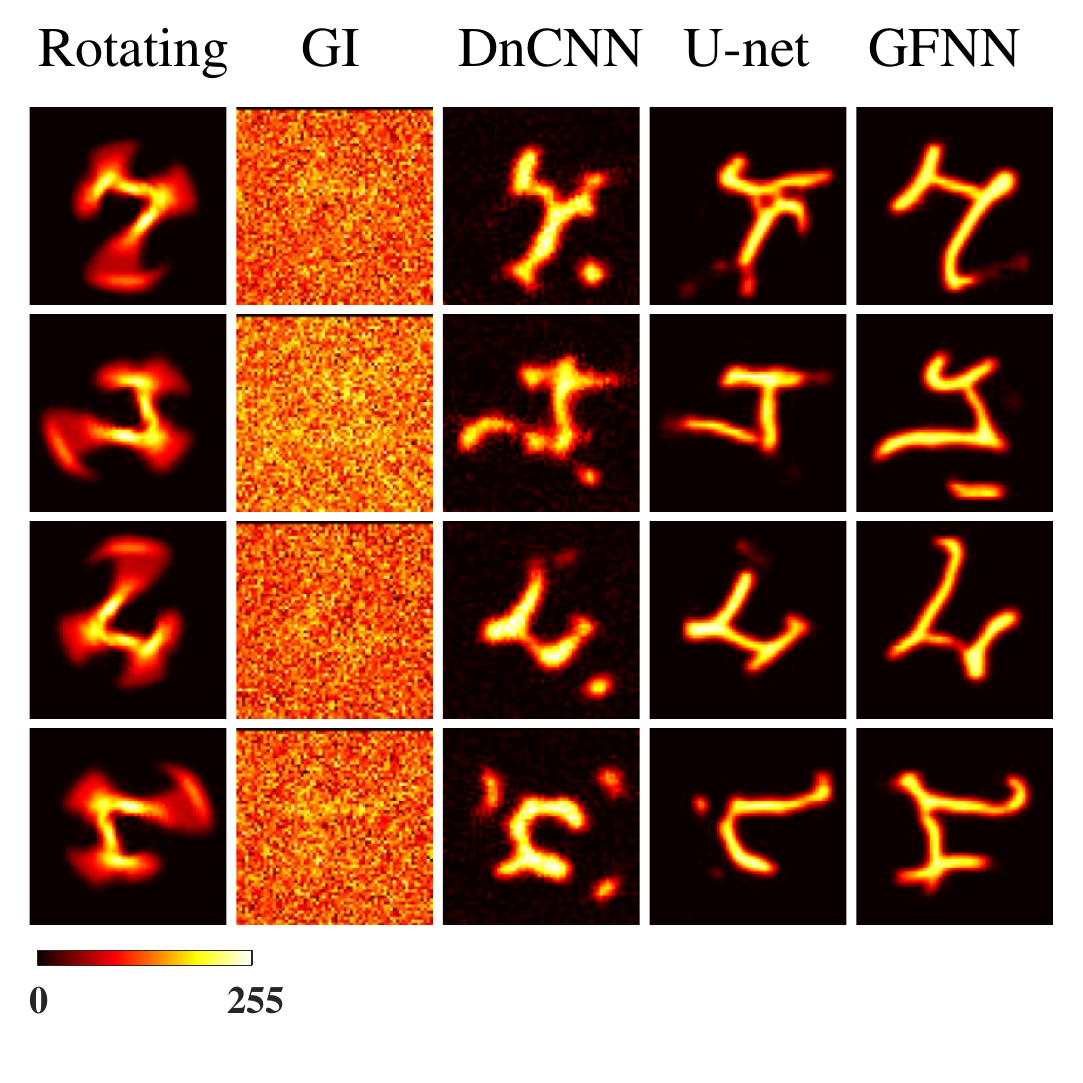}
\caption{The results of a high-speed moving object are reconstructed by different networks. The GI image has a sampling ratio of 3.125\%. The results trained by DnCNN and U-net are not satisfactory due to the blurred high-speed rotational images. GFNN eliminates the image blurring during the motion combined with the algorithm FMA and shows excellent test results. The number 7 is visible at all positions, close to the ground truth. In addition, the rotation angle of the detected objects here is three times of that in Fig. \ref{fig3}, thus the image exhibits slight distortion.}\label{fig5}
\end{figure}

Applying the network and FMA algorithm to the imaging process of moving objects. The number "7" is chosen for the simulation, with continuous samples of 300 frames. The rotation speed of the object is about $0.0375^{\circ}/ms$, the rotation angle is about $45 ^{\circ}$ during the acquisition, and the time consumed is about $1.2 s$. By setting the GF for each batch of 100 frames, the 300 consecutive frames are divided into 3 batches. The motion blur of the images is removed using the FMA algorithm and then input to the GFNN network for training. The final results are shown in Fig. \ref{fig5}.\par
\begin{figure}[htbp]
\centering
\includegraphics[width=0.62\linewidth]{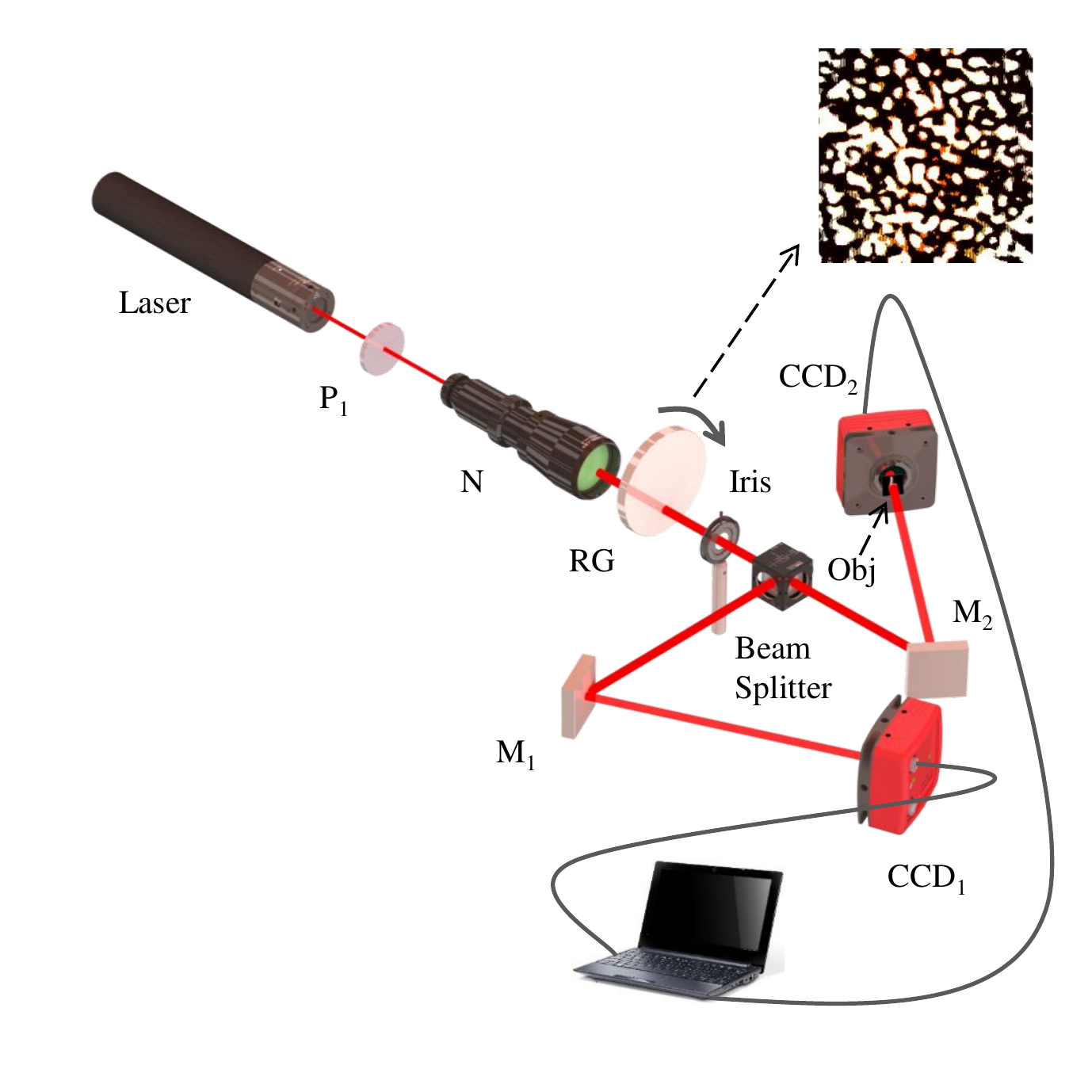}
\caption{Sketch of the experiment. Speckle patterns of the pseudo-thermal light collected during the experiment.}
\label{fig6}
\end{figure}

Simulations are completed under sufficient illumination, and the noise is mainly a statistical error. However, in the experiments with insufficient light, background noise or detection noise can also interfere with the output. The algorithm should fully consider these disturbances\cite{49}. To speed up the data processing under experimental conditions, we refer to the work of Wang F et al\cite{50}. We acquired the pseudo-thermal light speckle pattern and the object image, by entering them into the computer, and then correlated them. It is verified that it does not affect the results. The experimental setup is shown in Fig. \ref{fig6}.\par
The beam is generated from a $632.8 nm$ He–Ne laser (HNL225RB). The output power of the laser is around $2mW$. $P1$ is a polarizer to modulate the intensity of the laser beam. The light beam is expanded by a telescope (N, $\times 2$ magnification) and illuminates a rotating glass (RG) to produce a pseudo-thermal light beam. The beam via an Iris is then divided into two optical paths, an object beam, and a reference beam, by a 50:50 non-polarizing beam splitter. The pseudo-thermal light beam is collected, by two mirrors M1 and M2, to the Charge-coupled Devices (CCDs, MTV-1881EX), with a pixel size of $8.3\times 8.3 um^2$. And the distance from CCD to beam splitter is $30cm$. Two beams are connected by a data acquisition card in the computer and can be taken for correlation calculation. The sampling frequency in the experiment is 250 Hz. And the acquisition range in the experiment is $128\times 128$ pixels. And the rotation of the object is achieved by simulation.\par
\begin{figure}[htbp]
\centering
\includegraphics[width=0.8\linewidth]{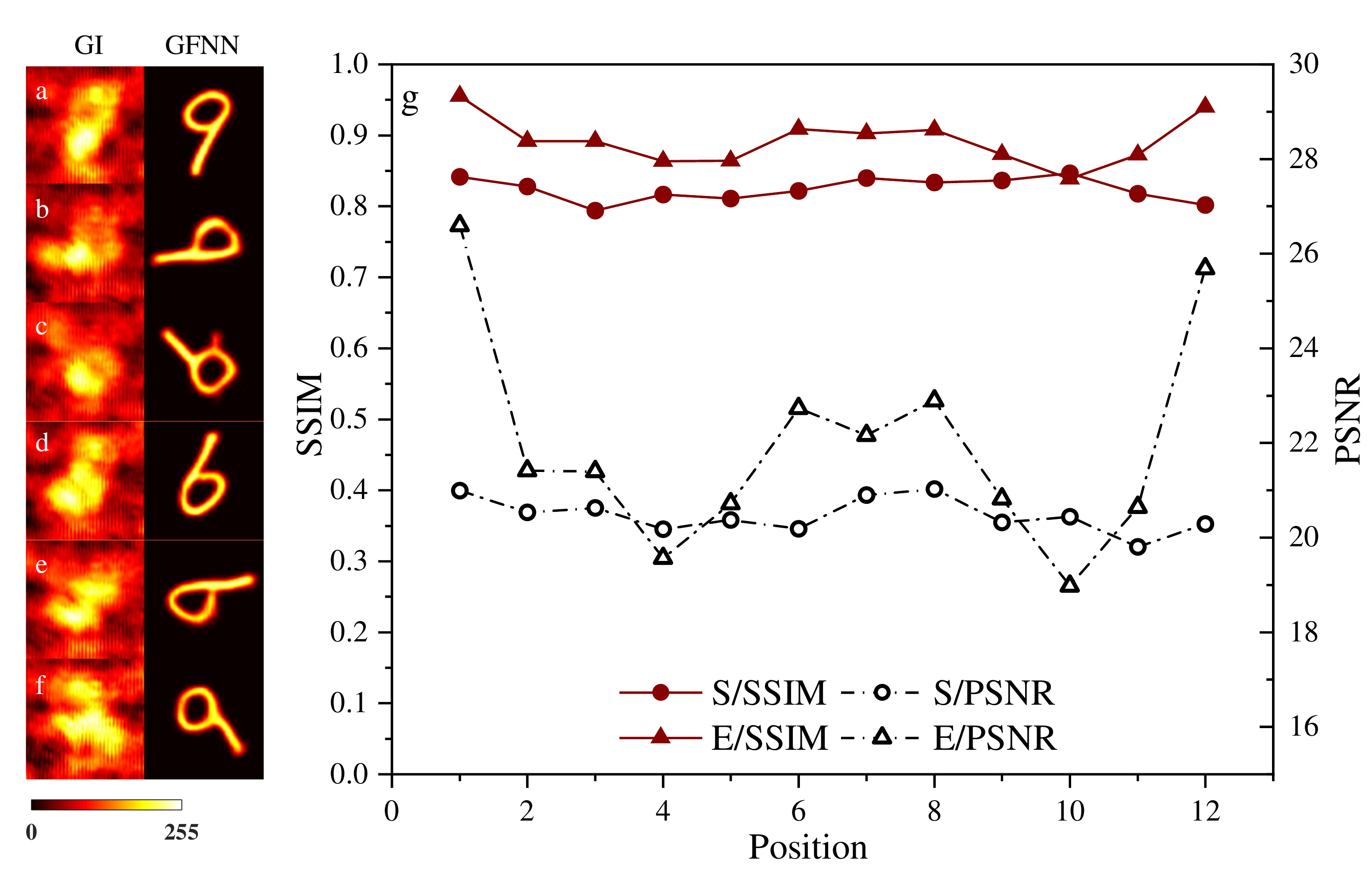}
\caption{The figure on the left shows the output of the network in the experimental environment. The image is clear with high quality. In c, the training result shows a slight distortion, which is owing to the turbulence of the experimental environment. The figure on the right shows the difference between SSIM and PSNR in the simulation and experimental results. (simulation: S, experiment: E)}
\label{fig7}
\end{figure}
Figure \ref{fig7} shows the output of the network in the experimental environment and the comparison between SSIM and PSNR of the experimental and simulation results. The experimental results appear fluctuations compared with the simulation results. The main reason is the effect of various parameters such as the intensity of light and the roughness distribution of the glass. The speckle patterns (Fig. \ref{fig6}) collected in the experiment are derived from pseudo-thermal light. In the simulation, we use randomly distributed speckle patterns and different sets of random speckle patterns among different BGFs to enhance the network generalization ability. It allows GFNN to perform equally well on training and test sets. But the computer randomly generated matrix values fluctuate widely, making the correlation process of ghost imaging disturbed. Therefore the simulation sacrifices some of the imaging quality while improving the applicability of the network. The average error of the experimental SSIM index and PSNR index compared to the simulation is 8.34\% and 7.26\%. The errors do not exceed 10\%. In addition, the bucket measurements are highly coincident in experiment and simulation, which also validates the concept of "learning from simulation" proposed by Wang F et al\cite{50}.\par
\section{Conclusion}
In summary, we propose a new multi-to-one network to improve the imaging speed and quality of ghost imaging at a low sampling ratio. The network exhibits excellent imaging results at a sampling ratio as low as 3.125\% and maintains a high generalization ability. For the ghost imaging of moving objects, the frame merging algorithm has been designed by combining the correlation characteristics between images. The algorithm eliminates the motion blur of high-speed objects and further improves the imaging quality of high-speed moving objects at a low sampling ratio. The detection noise and background noise in the experimental environment is also considered in this study. The results indicate that the relative errors between experiment and simulation are less than 10\%. Our purpose scenario will find potential applications in GI of moving objects. It can make it easier to image and track moving objects in large fields of view and the military target detection.\par

\begin{backmatter}
\bmsection{Funding}National Natural Science Foundation of China (No.12074350); Fundamental Research Funds for the Central Universities (No.2652018057); National College Students Innovation and Entrepreneurship Training Program (No.202211415093).

\bmsection{Acknowledgments}The authors thank Tong Bian for his helpful advice and discussions.

\bmsection{Disclosures}The authors declare no conflicts of interest.

\bmsection{Data availability} Data underlying the results presented in this paper are not publicly available at this time but may be obtained from the authors upon reasonable request.

\end{backmatter}

\end{document}